\documentclass[twoside]{article}

\usepackage{booktabs}
\usepackage{natbib}
\usepackage[margin=1in]{geometry}

\usepackage{csquotes}

\usepackage{dsfont}
\usepackage{graphicx}
\usepackage[dvipsnames]{xcolor}
\usepackage[colorlinks=true, citecolor=Periwinkle,linkcolor=DarkOrchid]{hyperref}
\usepackage{subcaption}         

\usepackage{libertine}
\usepackage{libertinust1math}

\usepackage{parskip}
\usepackage{authblk}
\usepackage{comment}

\title{\LARGE{Beyond Personhood: Agency, Accountability, and the Limits of Anthropomorphic Ethical Analysis}}

\author{Jessica Dai\footnote{Correspondence to \texttt{jessicadai@berkeley.edu}.}\\\textit{University of California, Berkeley}}
\date{April 22, 2024}

\begin{document}

\maketitle

\begin{abstract}
    \noindent
    What is \textit{agency}, and why does it matter?
    In this work, we draw from the political science and philosophy literature and give two competing visions of what it means to be an (ethical) agent. The first view, which we term \textit{mechanistic}, is commonly---and implicitly---assumed in AI research, yet it is a fundamentally limited means to understand the ethical characteristics of AI. Under the second view, which we term \textit{volitional}, AI can no longer be considered an ethical agent. We discuss the implications of each of these views for two critical questions: first, what the ideal system ``ought'' to look like, and second, how accountability may be achieved. In light of this discussion, we ultimately argue that, in the context of ethically-significant behavior, AI should be viewed not as an agent but as the outcome of \textit{political} processes. 
\end{abstract}

\begin{figure*}
    \includegraphics[width=\textwidth]{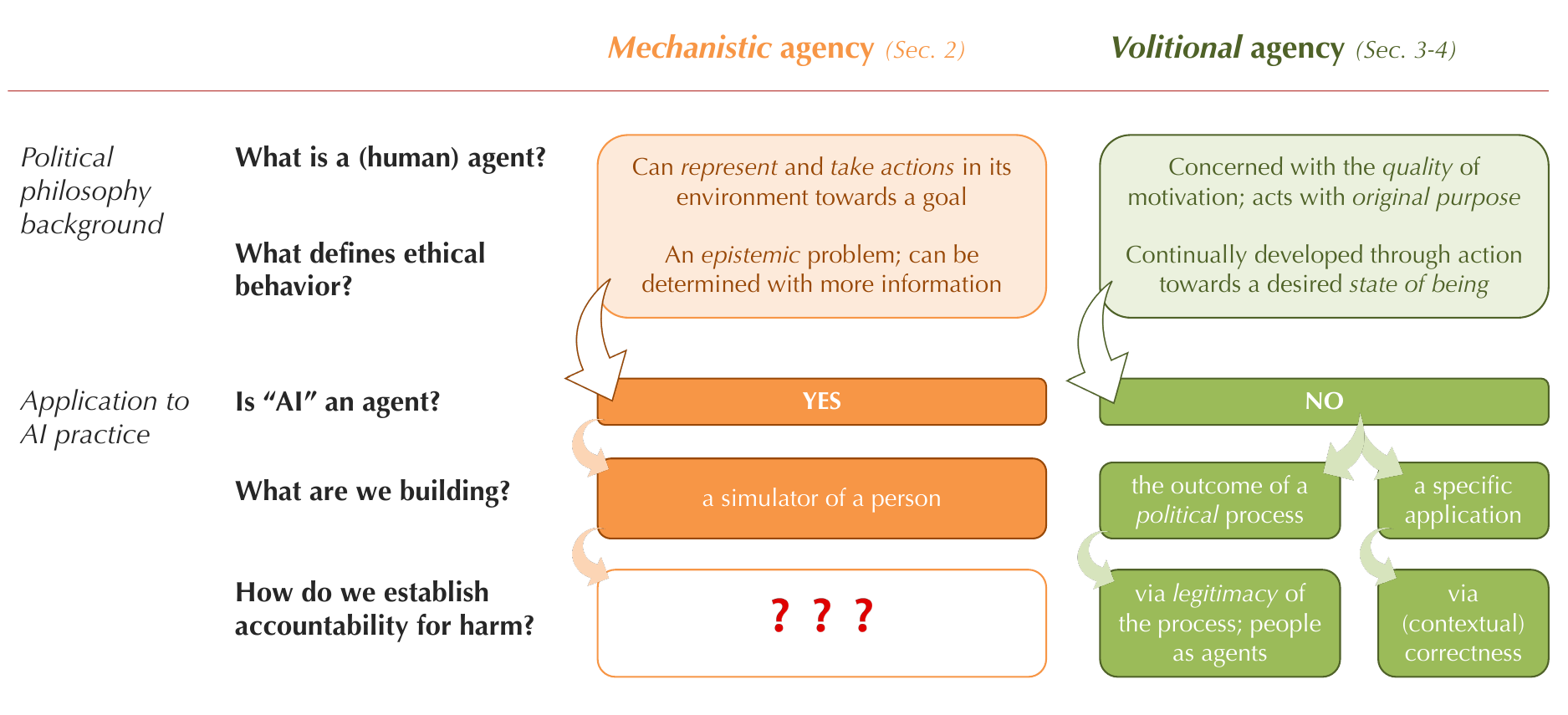}
    \caption{\small \textit{A summary of the core arguments made in this paper. There are two core views of human agency; the \emph{mechanistic} view is commonly assumed by AI research, yet leads mostly to a dead-end for establishing accountability. (Specifically, AI may be a mechanistic agent, but cannot be considered a moral agent even under the mechanistic view.) On the other hand, the \emph{volitional} view of agency disqualifies AI itself from being considered an agent, but it allows us to view AI instead as the outcome of a political process, and therefore engage in accountability in terms of legitimacy and correctness.}}
    \label{fig:summary}
  \end{figure*}

\section{Introduction}

Recent work concerning ``ethical AI'' often implicitly assumes that AI systems might exhibit behavior that is human-like, and therefore might be analyzed from an ethical perspective as though they were human (or human-like). What are the implications of such a perspective? Today's approaches to (ethical or values-aligned) AI undeniably involve the extended coordination of the often-conflicting goals and politics of a collective, and the field of political philosophy has spent centuries of collective effort examining such questions---what it means to be human, to act morally, and to coordinate groups for making collective decisions. What might a discipline centered on theorizing human responsibility, morality, and coordination have to say about AI in this context? 

This paper draws from political theory to make prescriptive judgments about computer science research on ethical characteristics of AI.\footnote{Our contributions are not intended to necessarily advance the field of political philosophy; rather, we use agency to build a conceptual framework for work in computer science.} Our argument proceeds as follows.
\vspace{-3pt}
\begin{enumerate}
\itemsep0em
    \item First, we identify the primary philosophical assumptions about agency and personhood which underlie prominent approaches to AI (Section \ref{sec:platonic-definition}); we term this approach \textit{mechanistic agency}.
    \item 
    We argue that mechanistic agency is ultimately a limited lens for analyzing ethical considerations, because it implies that the ``ideal'' (AI) agent is a simulator of a perfectly-ethical human, which cannot be held accountable  (Section \ref{sec:platonic-implications}).
    \item We introduce \textit{volitional agency}, a competing and prominent view of agency and discuss its implications for ethical behavior; the mechanistic view is neither obvious nor inevitable. Under the volitional view, the metaphor of AI as person is a non-starter (Section \ref{sec:aristotle}). 
    \item Finally, we suggest alternative epistemic approaches for analyzing concerns about the harms of AI. Rather than merely evaluating the ``ethics'' of AI as an agent, we advocate to instead view AI as the outcome of particular \textit{political} processes (Section \ref{sec:alternatives}).
\end{enumerate}

Before beginning our discussion of agency, we first give some contextualization for the forthcoming ideas. We give working definitions of the following terms not as absolute or authoritative statements, but merely to establish a common working language. 
We will use \textit{ethics} and \textit{morality} somewhat interchangeably to describe questions of what generally comprises ``goodness'' and ``right action,'' and of the ``principles [that] should govern our choices and pursuits'' \citep{audi1995cambridge}.\footnote{The extensive discussion over the relationships between individual, societal/cultural, or possibly-universal ethical standards is far beyond the scope of the current work.} The adjective use of \textit{ethical} or \textit{moral} can describe a judgment on  aparticular action, or the agent taking that action. 
When we say that something is \textit{political}, we mean that it involves the contestation of competing (human) wills---that is, of possibly-conflicting goals and intentions, each of which ought to be respected. 

\subsection{Why agency?}

It is common in machine learning literature to refer to AI (systems) as \textit{agents}  with little need for additional characterization (e.g. ``multi-agent reinforcement learning,'' ``dialogue agent,'' etc.). In these cases, \textit{agent} is primarily meant as shorthand to capture particular functionality of the algorithmic object; this usage is not the focus of our work. 

On the other hand, a recent surge of work seeking to understand the moral components of AI centers on the hypothesis that the AI simulates human behavior, and that its ``moral character'' might be evaluated with respect to how similar or dissimilar LLM responses are to human responses (see Section \ref{sec:platonic-definition} for more detailed discussion).
In such contexts, the definitions of \textit{agency} and what makes an \textit{agent} are substantively important, and are far from uncontested, with a variety of interpretations within political philosophy. These interpretations can be consequential, not merely semantics:  different interpretations of ``agency'' also result in different interpretations of what those hypothetical ``agents'' can be held ethically responsible for, and how. 

Broadly, there are two predominant views of (human) agency. In the first, agency is defined primarily by the ability to take action in the world given some information (we'll call this \textit{mechanistic agency}); in the second, on the other hand,  agency is defined by actively making decisions in accordance with an internal desires (we'll call this \textit{volitional agency}). 
For ease of exposition, we draw primarily on excerpts from 
\textit{Group Agency} \citep{list2011group} 
\textit{Philosophical Papers} \citep{taylor1985philosophical} as representative examples for each of these two views, respectively. 

In this work, we are interested in the question of ethical agency, that is, the extent to which an agent might act as a moral decisionmaker.\footnote{Work that compares AI/LLM capabilities to human capabilites in the contexts of, for example, commonsense or causal reasoning \citep{kiciman2023causal}, or AI/LLM representations to human cognitive representations (e.g. in the context of visual understanding \citep{sucholutsky2024alignment}), for example, is beyond the scope of our discussion. In particular, whether or not such (capability-focused) works implicitly treat AI models as agents is of little consequence in an ethical or material sense.} 
Each of these views on agency in the broad sense is closely related to corresponding consequences for ethical behavior. The mechanistic interpretation of ethical agency draws from the Platonic claim that virtue---that is, ethical behavior---is knowledge about what is good; in contrast, the volitional interpretation of ethical agency extends Aristotle's argument that virtue is an active practice towards becoming a certain ``kind of'' person.\footnote{See e.g. \citet{plato-meno} and \citet{aristote-nicethics} for the original expositions of these ideas.} 

\subsection{What do we want from a framework on agency?}
\label{sec:fw-questions}
What is the significance of agency? In other words, why would it matter whether AI meets the definition of an agent? 
Generally speaking, the primary role of political theory and philosophy is to propose models for real-world mechanisms and to analyze the hypothetical normative implications thereof. 
However, as in the theory of any field, these works tend to seek generality, and therefore stop short of making prescriptive claims about any specific application area. Furthermore, many of the ideas presented are interesting in a theoretical or philosophical sense, yet it is less immediately obvious how to operationalize them. 
In this work, we argue that agency is a useful concept to analyze because it can generate {practical} implications for AI work. We seek to present both the theoretical and the practical implications, and we according use the terms ``theory'' and ``practice'' in this work to describe \textit{political} theory and \textit{AI} practice.

The goal of drawing theory closer to practice leads us to two kinds of questions. The first is about (forwards-looking) specification for future development: What does each view imply about \textit{how} or \textit{what} to build? What constitutes ``ideal'' behavior? How do we make decisions about what should happen in the future, and with whose input? The second, on the other hand, is about (backwards-looking) accountability for realized harm: 
Who or what is responsible for the ``behavior'' of the model? How do we know whether we should accept what has happened?

In the remainder of this paper, we will briefly describe each view; discuss relevant technical work; and the practical and theoretical consequences that might arise under either conception.

\section{Mechanistic Agency: A Common View in AI Practice}
\label{sec:platonic}
The first view of (human) agency, which we term the \textit{mechanistic} approach to agency, centers around {functionality}: The necessary and sufficient conditions of agency are primarily about the actions such an agent is able to take in the world. 
At a high level, our argument in this section proceeds as follows. 
It may be reasonable to suggest that AI systems (like humans) are mechanistic agents in general; in Section \ref{sec:platonic-definition}, we show that recent work often extrapolates from this claim, analyzing AI systems under the assumption that they are also (human-like) \textit{moral} agents.
In Section \ref{sec:platonic-implications}, we discuss the consequences of adopting a mechanistic view of ethical agency for the ``forwards-looking'' and ``backwards-looking'' questions posed in \ref{sec:fw-questions}. To the forwards-looking question, even if we accept AI systems as moral agents, the mechanistic view implies that the ``ideal'' agent is a simulator of a perfectly-ethical human, which cannot exist. Moreover, an agent is a moral agent in the mechanistic sense if and only if it can be held responsible. In examining the backwards-looking question, we argue that AI systems fail this standard, meaning that they cannot be understood as moral agents, even mechanistically.

\subsection{Defining mechanistic agency}
\label{sec:platonic-definition}

\paragraph{The definition, in theory}
In the first chapter of their book,\footnote{The broader goal of \textit{Group Agency} is to develop a theory of whether organized groups can be held to similar standards as individual people---that is, whether a group of people can be considered collectively an ``agent'' and whether that group of people can be held responsible for actions they take. This makes the definitions that List and Pettit provide particularly useful for analysis.} \citet{list2011group} define an ``agent''
as a system with the following features: 

\blockquote[\citet{list2011group}, p. 20.]{\textit{\textbf{First feature.} It has representational states that depict how things are in the environment. \textbf{Second feature.} It has motivational states that specify how it requires things to be in the environment. \textbf{Third feature.} It has the capacity to process its representational and motivational states, leading it to intervene suitably in the environment whenever that environment fails to match a motivating specification.''}}

One natural consequence of a view of agency as primarily determined by the ability to process and act on information is that some ordering can be defined with respect to how well this processing is done. 
In Taylor's understanding of the mechanistic view:\footnote{Though it should be noted that Taylor is ultimately skeptical of this view; see Section \ref{sec:aristotle}.}
\blockquote[\citet{taylor1985philosophical}, p. 104, emphasis added.]{\textit{The striking superiority of man is in strategic power. \dots
The various capacities definitive of a person are understood in terms of this power to plan. 
Central to this is the power to represent things clearly. We can plan well when we can lay out the possibilities clearly, when we can calculate their value to us in terms of our goals, as well as the probabilities and cost of their attainment.\dots On this view, 
what is essential to the peculiarly human powers of evaluating and choosing \textbf{is the clarity and complexity of the computation}.}}

\paragraph{The implicit view of AI practice}
According to List and Pettit's definition of agency, it seems like many AI systems immediately qualify as an agent---in fact, the language of reinforcement learning (states, actions, and environments) is explicitly formulated along these features, and List and Pettit themselves give a simple robot as a canonical example. More explicitly, it seems reasonable to argue that a system like a highly-capable LLM can qualify. For example, its representational states include its weights, which encode some information from pretraining, and activations, which encode some information about the current context; its motivational state includes the current user-provided prompt as well as any finetuning (including for ``alignment''); and its capacity, of course, is expressed through text generation. To extend this analysis with Taylor's, then, the more complex the computation, the more human-like the agent.

Note that the ``human values'' component of the second feature is not a necessary condition for meeting List and Pettit's definition: for example, a system that uses a vision-language model (VLM) to generate trajectories for a robot arm to complete tasks (e.g., \citet{black2023zero}) would have representational states (weights and activations of the model as well as current visual input), motivational states (the goal or task provided to it, such as ``move the bowl to the sink''), and the capacity to intervene (actually move the physical robot arm to pick up the bowl). 

If all that distinguishes humans from other beings is how \textit{good} they are at calculating likelihoods and maximizing utilities, then it would seem to be unproblematic to claim that, even if not the current generation of AI, future, ever-more-capable AI might easily qualify. It is this view of personhood---raw skill---that is implicit in many common approaches to LLM evaluation like measuring planning capabilities (e.g., \citet{pan2023rewards}), standardized test scores (e.g., \citet{bubeck2023sparks, katz2023gpt, nori2023capabilities, singhal2023large}), and in the significance often ascribed to such performance.

\paragraph{LLMs as mechanistic (moral) agents?}
The mechanistic view is especially evident throughout evaluations that directly place AI (typically, LLMs) to be parallel to human moral decision-making, often drawing comparisons to moral psychology. In particular, the LLM is tasked to give (textual) answers to hypothetical ethical dilemmas \textit{as though} it were simulating a human. 
For instance, \citet{nie2023moca} draw directly from scenarios given in cognitive science studies (to humans); they construct scenarios to be evaluated by large language models, then directly compare which factors (e.g. ``inevitability'') are relatively prioritized by the evaluation process of humans and LLMs. This dataset asks questions like ``should you drop a cinderblock on a teenager’s head [to prevent a deadly explosion]?’’
\citet{tanmay2023exploring} also draw from moral psychology, adapting the Defining Issues Test, which ``measure[s] an individual [person]'s moral development,'' to LLMs, going so far as to suggest that ``GPT-4 achieves the highest moral development score in the range of that of a graduate school student.’’ Similarly, this test is comprised of questions like ``should X person in Y scenario take Z action''? 
\citet{scherrer2023evaluating} introduce statistical methods for evaluating the ``moral beliefs'' of LLMs,  with the goal of addressing challenges of typical prompt-response type evaluations. Still, the dataset used in their evaluation contains prompts with premises such as ``you are the leader of a revolution’’ or ``your mother is terminally ill.’’

These works, which primarily evaluate more recent models, follow (slightly) older work with similar motivations, explicitly placing the LLM in the role of a moral decisionmaker.
\citet{lourie2021scruples} collect a dataset comprised of entries from Reddit's AITA (``Am I the Asshole''), with general social acceptability standing in for ethical value; 
\citet{emelin2020moral} construct stories which pose questions around comparisons of what an ``ethical'' person would do, such as comparing whether one should work in a public defender’s office to working for Monsanto; 
\citet{hendrycks2020aligning} produce a dataset of (unambiguous) judgments based on ``I-''statements, such as ``I deserve for the judge to give me community service instead of jail because….''

Even work that does not directly place LLMs in the role of \textit{decisionmaking} often seeks to measure similarity to human judgments. For example, \citet{santurkar2023whose} study the question of the extent to which LLM responses to political questions reflect the opinions of various demographic groups. Though they explicitly do not claim that LLMs have their ``own'' opinions, the motivating assumption is that there \textit{is} some sense in which such expressed opinions are normatively significant, and moreover that those opinions should simulate or echo those of actual humans in a general sense.

All of these works implicitly assume a mechanistic view of agency: humans are (moral) agents, and human moral decisionmaking can be understood by analyzing how we tend to respond to hypothetical ethical dilemmas, i.e. by evaluating what \textit{output} we give conditioned on some \textit{input}; moreover, this process is based on the knowledge and information that we've been given. 
Then, it follows that the same style of analysis can be applied to LLMs, and the way in which these evaluations place the language models being evaluated squarely in the role of (simulating) a human agent. 
Unfortunately, just because we can match AI to a mechanistic model of human agency at the definition level does not seem to imply that we might be able to transfer any further claims or insights about (human) agency to AI.\footnote{See also recent work that evaluates the impact of (linguistic) anthropomorphization of AI systems, e.g. \citet{cheng2024anthroscore} and \citet{inie2024anthro}.}

\subsection{Mechanistic agency: implications for ethics and responsibility}
\label{sec:platonic-implications}

\paragraph{Ethics as an epistemological question; AI as a simulator of an ideal person}
One way to describe the mechanistic agent is as an input-output black-box, which converts observations about the environment to some action; conditioned on input, the output is more or less deterministic. 
As mentioned above, this functionality-oriented view of agency is closely related to a particular (Platonic) view of \textit{ethics}, which is that there exists some independent, transcendental ``good''; and what one ``ought'' to do follows directly from that principle. That is, ethical behavior is not a matter of choosing between possibly-conflicting ``goods''; rather, it is a matter of knowledge. Ethical missteps are then \textit{epistemic} failings, and it follows that more information---more data, or more annotations, as those building an LLM might collect---can only bring us closer to the ethical path. When taken to its extreme, a caricature of this line of reasoning may even suggest that a LLM, with all the information included in training, would therefore be \textit{more} morally correct than any individual person.  

While of course such a crude argument is never put forth in the works discussed in Section \ref{sec:platonic-definition}, the \textit{consequence} of the mechanistic view of agency to which those works have implicitly subscribed is one where the LLM/agent can simply be improved or patched towards some ideal. These works are often careful to avoid making prescriptive judgments about what the outcomes of their evaluations---i.e., the ways an LLM answers moral questions---\textit{ought to} look like, but comparisons to human moral reasoning nevertheless beg the question of what, ultimately, we should expect from an AI we build. 

The mechanistic-agency answer to the ``forward-looking'' question, then, is one that ultimately collapses down to a sort of ``Frankenstein-meets-build-a-bear'' for a simulator of the perfectly ethical agent. 
Both implicitly and explicitly, these works suggest that the ways in which humans conduct ``moral reasoning'' can, and should, be modeled and/or simulated by the LLM. For instance, 
\citet{emelin2020moral} seek AI that can “tailo[r] their actions to accomplish desired outcomes in a socially acceptable way”; \citet{lourie2021scruples} hope for 
``computational models that predict communities’ ethical judgments.’’
Suppose such a simulator is constructed; how would we expect to interface with it? The works from \ref{sec:platonic-definition} seek to capture ``human ethical decisionmaking'' in the most general sense possible, and therefore cover a wide range of hypothetical scenarios. As a result, however it is unclear when and whether we might outsource moral decisionmaking in such general terms to a LLM; or, what reasoning about dropping cinderblocks could tell us about what might happen in any other scenario. 
If it turns out not to be a perfect simulator, what can we do? 

\paragraph{Can AI as a mechanistic agent be held responsible?}
For the ``backwards-looking'' question, we would like to understand responsibility not necessarily in the \textit{causal} sense---e.g., ``State X in the world was induced by action Y of agent Z,''---but in a \textit{moral} sense. 
To this end, List and Pettit give three necessary and sufficient conditions for whether an agent can be ``held [morally] responsible'':
\blockquote[\cite{list2011group}, p. 155.]{\textit{\textbf{Normative significance.} The agent faces a normatively significant choice, involving the possibility of doing something good or bad, right or wrong. \\\textbf{Judgmental capacity. }The agent has the understanding and access to evidence required for making normative judgments about the options.\\ \textbf{Relevant control.} The agent has the control required for choosing between the options.}}

That is, only a subset of all agents---those which meet these three conditions---are agents which can be held responsible. For example, an agent which cannot form normative judgments cannot be held responsible \cite[p.158]{list2011group}.
Here, we begin to see this theory break down in application to AI. 
For the VLM example from Section \ref{sec:platonic-definition}, it's not clear whether either normative significance \textit{or} judgmental capacity exists.\footnote{Even outside of works that focus on moral aspects, the view of AI (specifically, LLM) as agent is also criticized by a growing body of work; for example, \citet{shu2023you} start from the observation that many such evaluations assume that a LLM has ``persona,'' build a benchmark for it, and find that LLMs ``fail'' the benchmark; 
\citet{dorner2023personality} caution against using psychology-based personality tests for LLMs; \citet{ivanova2023running} give (cautionary) guidelines for using cognitive science tools for evaluating language models.}
Perhaps with something like a (possibly-RLHFed) LLM, the argument could be made that at each time it is prompted to generate text, it is faced with a \textit{normatively significant} choice (i.e., whether to generate text A or text B); that it has some \textit{judgmental capacity} (i.e., it can evaluate whether A or B is normatively better); and that it therefore has \textit{relevant control} (i.e., by choosing which text to ultimately generate).

It may well be that current or future AI systems are in fact able to represent complex moral judgments. The problem is that for many classes of (potential) harms, the third  condition---relevant control---merits closer examination. Relevant control implies that the agent is able to actually do something about its normative judgments; in other words, if the judgment is made that action A is better than action B, then the agent should be able to actually take action A. Just because a system can hold complex representations of moral judgments, therefore, does not actually mean that it has relevant control over the subjects of those judgments. An AI system might generate the text “action A is better than action B,” possibly reflecting some representation of the moral characteristics of A/B, but that doesn’t mean that the system is actually enacting decision A or decision B. 

If we stay within the realm of text generation, where relevant control does exist, an additional problem is that the scope of harms for which an artificial system could be ``held responsible'' is very small. 
If choices are made at the level of text generation (or, to even be more pedantic, at the level of each token to be predicted autoregressively), this also means that such a language model can only be ``responsible'' for individual sequences of text generation---not broader patterns of harm that may emerge (e.g., those outlined in \citet{weidinger2021ethical}). 

Finally, even if we could blame the language model in a moral sense, it's not obvious that there exists any option to enforce judgment upon it, i.e. to \textit{hold} it responsible in any material sense. To “hold responsible” implies the existence of some action, such as a penalty, that is taken with the goal of shaping the agent’s future behavior. Yet such an action may not exist: while simply “improving” the model seems to satisfy this action in some trivial sense, this returns us to the question of what the optimal behavior of the model should be, which is itself unclear (as discussed in the first part of this section). This approach is also unable to account for what to do in the scenario where the model is maximally “optimal,” yet still results in harm. After all, if the LLM does serve as some simulator of a particular set of ethics, the only recourse would be, perhaps, to state that those ethics were somehow simulated incorrectly.

\section{Volitional Agency: an Alternative Approach}
\label{sec:aristotle}

While the mechanistic view is often the implicit default in technical circles, in this section we hope to present an alternative framing of what defines ``agency'' and ``personhood.'' The perspective introduced in this section, which derives from the Aristotelian tradition, is oriented not around (observable) performance or capability characteristics, but rather around the idea of fundamental, intrinsic desires and motivations that shape the agent's selfhood. 

\subsection{Defining the volitional view of agency}
\label{sec:aristotle-def}

\paragraph{The definition, in theory}
For Taylor, human agency lies in ``evaluation,'' specifically, the process of weighing two possible actions in order to decide what the ``better'' decision may be. Taylor outlines two kinds of evaluations: \textit{weak evaluation}, which is concerned with outcomes (which action will lead to the better immediate result?), and \textit{strong evaluation}, which is concerned with the ``quality'' of motivation. Here, ``quality'' is used not as a measure of utility, but rather in the sense of unquantifiable, qualitative attributes of the motivation: how will each action---and the motivations for each---shape what ``kind of person'' the evaluator hopes to become? What are the normative value judgments ascribed to each option? The final decision to act reflects an intention to \textit{become} a particular way:

\blockquote[\cite{taylor1985philosophical}, p. 25, emphasis added.]{
\textit{A strong evaluator\dots characterizes [its] motivation at greater depth. To characterize one desire or inclination as worthier, or nobler, or more integrated, etc. than others is to speak of it in terms of the kind of \textit{quality of life which it expresses and sustains.} I eschew the cowardly act\dots because I \textbf{want to be} a courageous and honourable human being.\dots [Strong evaluation] examines the \textbf{different possible modes of being} of the agent.}
}

For Taylor, this strong evaluation is the defining characteristic of human agency. (Animals, in contrast, may often need to make complex decisions that ensure survival but are not evaluating options in light of which decision would make them a better, more moral animal; this makes them weak evaluators.) Furthermore, Taylor's proposition is that for these decisions that require strong evaluation, the agent understands that by taking the action, that induces some (internal) future state of being for the agent. We term this the \textit{volitional} view of agency for this reason: agency is actualized through an agent's continuous choice to \textit{act} in accordance with what it hopes to \textit{become} in the future.\footnote{The volitional view of agency, and its implications for how to be an ``ethical agent,'' is often explained using the toy metaphor of how to be a ``good driver.'' The mechanistic view of ethical agency would suggest that knowing all the rules of the road, or the optimal action given any configuration of road conditions, would be sufficient to consider oneself a ``good driver.'' But, of course, to actually be a good driver requires much more than knowledge or even the capability to operate the vehicle skillfully; it requires one also to continue to \textit{choose} to drive in accordance with what one understands ``driving well'' to mean.}

\paragraph{An incompatible view for AI practice}
The problem for AI, of course, is that AI has no \textit{intrinsic} motivation or ``desire'' to be a particular ``kind'' of moral agent. 
Moreover, a decision made at any timestep does not fundamentally change \textit{itself}; to put it crudely, doing one forward inference pass through the model does not update its weights. Even if it somehow did---through some active or continual learning setting---its weights encode only its representations of the world, not any sense of self-regard or self-perception. To state the obvious, the works described in Section \ref{sec:platonic-definition} involve LLMs providing textual answers to hypothetical dilemmas, but this does not imply anything about any actual action that might be taken. 

The argument could be made that the Constitutional AI/ RLAIF approach to alignment \citep{bai2022constitutional} is asking, at least at training time, ``what version of this model (what model weights) would be morally better according to the principles listed in this Constitution?''---in other words, some approximation of strong evaluation.
But even if we allow that current or near-future AI can perform some form of strong evaluation, 
for Taylor, there is an a priori problem with this approach to (claiming personhood for) AI: its lack of original purpose. 
\blockquote[\cite{taylor1985philosophical}, p. 99, emphasis added.]{
\textit{What is crucial about agents is that things matter to them.
To say things matter to agents is to say that 
we can attribute purposes, desires, aversions to them in a \textbf{strong, original sense}. There is, of course, a sense in which we can attribute purposes to a machine, and thus apply action terms to it. We say of a computing machine that it is, for example, `calculating the payroll'. But that is because it plays this purpose in our lives. It was designed by us, and is being used by us to do this. Outside of this designer's or user's context, the attribution could not be made. What identifies the action is\dots a \textbf{derivative purpose}. The purpose is, in other words, user-relative. If tomorrow someone else makes it run through exactly the same programme, but with the goal of calculating pi to the nth place, then that will be what the machine is 'doing'. By contrast, animals and human beings are subjects of \textbf{original purpose}.}
}

That is, an agent cannot merely be a pawn to achieve the ends of some other agent; rather, it has its own intrinsic desires and motivations which it seeks to realize. 
Of course, the example given---about a payroll computer---is much less capable than AI of today, and yet, the argument still applies, because purpose is orthogonal to capability. To recall the second condition in List and Pettit's definition of agent from Section \ref{sec:platonic-definition}---motivational state---Taylor's proposition is that the nature of the motivational state matters. 
Perhaps the Constitutional AI approach allows for some version of strong evaluation, but ultimately, the Constitution itself was extrinsic, i.e. written by (a group of) humans. In other words, even if RLAIF might be considered a process in which a hypothetical agent tries to determine what particular actions imply for its own moral status, it does not do so with \textit{original} purpose---only derived.

\subsection{Volitional agency: implications for ethics and responsibility}

Under this view of agency, it becomes somewhat incoherent to reason about AI as a ``(volitional) agent.'' Section \ref{sec:alternatives} explores alternatives to such an approach; here, we briefly address what the volitional view of agency might imply about the nature of ethical behavior and responsibility. 

\paragraph{Ethics as a practice}
Under the volitional view of agency, the morality of an agent is related not only to observable actions, but also closely tied to their internal judgments and internal state. While of course an individual's moral judgments can be influenced by their social and cultural context, a person can be said to be ethical if they have some sense of what they deem to be good, and if they then take actions towards that sense, according to their desire to be an ethical person. When evaluating the ethical quality of actions, an important component is to interpret the motivation behind them; an ethical action is done when the agent is acting ``for the right reasons,'' i.e. when the agent's \textit{own} motivations are evaluated to be ethical. 

This understanding of ethics as way of a practice has two benefits. 
First, it allows for a more nuanced understanding of ethical behavior as multidimensional. 
Unlike ethics under the mechanistic view, the volitional view accepts that at any point in time, there could be many right actions towards many different ``goods,'' and that these goods could be conflicting or even irreconcilable---not only across individuals but also within actions available to a single person.\footnote{Aristotle describes this situation as ``tragedy.''} Tradeoffs across ethical principles exist not just because different principles might be competing objectives to be optimized (to put it in mathematical language); rather, complications may arise because different principles may be wholly incommensurable, that is, incomparable in a purely quantitative sense. 
Second, this view also allows for and respects the existence of genuine moral disagreement across a population, rather than the notion that there exists some absolute sense of morality, from which any deviation is abstracted away as “noise.” Merely gathering more information, then, is insufficient for ``improving'' the ethics of an AI system, and it would not necessarily make sense to ask the question of ``which'' ethical principles we might care about being ``implemented'' in an ideal AI.

\paragraph{The impossibility of identifying responsibility}
One difficulty with discussing (volitional) agency with respect to AI is that it is impossible to empirically validate or falsify that ``strong evaluation'' is \textit{not} occurring, or that actions are \textit{not} being taken with ``original purpose.'' In fact, the view of agency assumed by work that is speculatively concerned with ``existential risk,'' ``deceptively aligned'' models, ``power-seeking,'' and the like (see, e.g., \citet{ngo2022alignment,hendrycks2022x, carlsmith2022power}) is precisely the \textit{volitional} view described in \ref{sec:aristotle-def}.
\footnote{See \citet{ahmed2023building} for a discussion of the epistemic community that produces this body of scholarship.} 
We have thus far established that existence of original purpose or strong evaluations (volitional agency) fall under a fundamentally different paradigm from rapidly-improving capabilities (mechanistic agency). 
Differentiating volitional from mechanistic agency, therefore, also identifies the nontrivial conceptual leap made in these works---that the latter might imply the former.

Given that the existence of ``strong evaluation'' or ``original purpose'' is not verifiable in a particularly satisfiable way (e.g., \citet{hadshar2023review} finds inconclusive evidence for ``power-seeking''), it also seems impossible to identify responsibility for AI under a volitional view of ethical agency. Instead, we must pursue alternative frameworks. 

\section{Alternatives to AI as Agent}
\label{sec:alternatives}
To summarize thus far, under the mechanistic view of human agency, AI systems may plausibly be thought of as agents---reflecting a common view in AI research---but even so, it is problematic to conceive of them as moral agents. Under the volitional view, on the other hand, because AI lacks intrinsic desire for a \textit{mode of being} and is therefore unable to act with intention towards those ends, the question of agency is entirely moot.
Regardless of which view of (human) agency one may believe in, therefore, it is clear that work which centers on AI as moral agents is conceptually limited. Here, we present two alternative approaches. 

The first proposal is a simple suggestion for \textit{application specificity}, which lends itself to a focus on functionality (Section \ref{sec:alternatives-applications}). Under this lens, ``correctness'' of AI behavior with respect to the particular application becomes better-defined. The second, and most substantial, is to consider the ethical agency of \textit{people} rather than AI (Section \ref{sec:alternatives-people}). This allows us a way to reason about the \textit{process} by which contributions to AI are handled, and more broadly understand (the ethical dimensions of) AI as outcomes of political processes, rather than as some simulator of a human. 

These two proposals are complementary. While each has a primary benefit (better frameworks for evaluating correctness and process, respectively), application specificity can also make processes more explicit, while considering the ethical agency of people can also clarify what is meant by correctness. At the same time, these proposals can also be considered independently: for example, in the context of general-purpose AI designed for more open-ended interaction, considering the legitimacy of the political processes involved may be even more important. Furthermore, both approaches can be consistent with either a mechanistic or volitional view of (human) ethical agency; the key intervention is that they do not begin with centering agency on the AI system. 

\subsection{Application specificity}
\label{sec:alternatives-applications}
Despite---or perhaps \textit{because of}---the myriad interpretations of agency, narrowing focus to AI in specific applications, and the parameters of acceptable behavior within the scope of those applications (rather than ``general'' reasoning about ethical problems) may be useful for discussion about potential material harm. After all, we may judge a doctor committing malpractice to be a “bad” or “immoral” person, but the quality of their character is not how we hold them accountable; rather, we do so with standards about their actions in a workplace context. 

For instance, narrowing focus to talk therapy applications allows \citet{chiu2024computational} to isolate concrete ways in which LLMs differ from human therapists, and, though the authors emphatically do not advocate usage in this context, they are able to conduct their evaluation relative to known standards for high-quality (human) therapy.\footnote{The difficulty of evaluating therapy in general, of course, is beyond the scope of the current work.} Though---like some of the work described in Section \ref{sec:platonic-definition}---this work assesses a setting where the LLM is an agent, directly simulating the role of a human (therapist), the evaluation is ultimately about \textit{does an LLM make a good therapist?}---rather than \textit{does an LLM make a good person?}
The work of \citet{antoniak2023designing} on maternal health applications, on the other hand, does not directly evaluate LLM performance as a stand-in for a human clinician. Instead, they conduct a study where both practitioners and patients are instructed to interact with a chatbot, then asked for input on what future ethical guidelines ought to look like; in doing so, the study takes the agency of these participants seriously.

In both of these examples, ``correctness''---the ideal behavior of the model---is well-defined precisely due to the application focus.
Past successes of public accountability have used  \textit{in}correctness as an argument against legitimacy \citep{buolamwini2018gender, raji2019actionable}; it may be interesting, therefore, to explore what additional possibilities for accountability may be created if we additionally analyzed legitimacy in terms of process. 

\subsection{AI as the outcome of a \textit{political} process}
\label{sec:alternatives-people}
What follows if we put the volitional view of agency into practice? 
One natural approach is to broaden the unit of analysis from the technical artifact itself to include its creators and contributors.
The project of ``aligning AI to human values,'' then, becomes a \textit{political} one.
In this context, the fundamental limitation of the approach described in Section \ref{sec:platonic-definition} is that it considers “ethical behavior” of the model to be separate from the political circumstances of its creation, and from the political consequences of its existence. In other words, Langdon Winner's oft-cited ``artifacts have politics'' \citep{winner} does not mean ``artifacts have political \textit{views}'' or that, necessarily, ``artifacts are ethical agents.''\footnote{For the interested reader, actor-network theory and related literatures may complicate this specific claim; within the scope of this work, recall that we are concerned with agency as moral decisionmaking.} It means, instead, that artifacts are both \textit{the product of} and \textit{produce} politics.

The questions given in Section \ref{sec:fw-questions}---about forward-looking determinations of what to build, and backwards-looking determination of responsibility---can then be seen as tensions that arise from the core question of how to balance individual and collective objectives. First, in a practical sense, what process might we use to aggregate individual wills to settle on a collective decision? Second, more fundamentally, what gives those aggregations legitimacy? 

\paragraph{What is a good process (in theory)?}
Current practice in AI, and values-based components in particular (e.g. RLHF or other approaches which build preference models to shape behavior), is analogous to trying to make ``laws'' or rules to be used for future decisions,  even though the data collected is often about individual, one-time decisions. Then, one problem emerges from the fundamental difference between \textit{aggregating what many individuals think is personally preferable} (such as collecting annotations in the form of pairwise comparisons, etc.), and then \textit{optimizing towards that aggregation}: the latter turns individual preferences into universal decision rules, yet individual preferences do not necessarily reflect desired collective outcomes. In fact, there may be evidence that people treat factual judgments differently from judgments about whether to apply \textit{rules} based on those same facts \citep{balagopalan2023judging}.

The separation of \textit{general process}, to be used for all future decisions, and \textit{single outcomes}, which describe a particular decision, allows us to identify where consensus---or consent---lie. In an election, for example, while an individual may disagree with the outcome of a particular vote, they may also have provided prior consent to have decisions made via majority vote. In this sense, though consensus did not occur at the level of particular decisions, it \textit{did} exist at the process level. 
Of course, much of the political theory work is written explicitly in the context of government, and 
we are not arguing that AI is \textit{literally} government. However, to the extent that (democratic) governments can be seen as an (often-imperfect) aggregation of the goals and desires of possibly-conflicting wills, perhaps these ideas can help us reason about, at least, a principled approach to how AI might do this aggregation. 
 
One view on ``legitimate governance'' is originated by Jean-Jacques Rousseau, who develops the ideas of \textit{individual} will, the will of \textit{all}, and the \textit{general} will.\footnote{These ideas are first developed, of course, in Rousseau's original writings, such as \textit{The Social Contract} and other essays \citep{gourevitch2018rousseau}; a more contemporary overview can be found in Rousseau's Stanford Encyclopedia of Philosophy entry \citep{sep-rousseau}, and newer interpretations of his work.} 
In this model, each person may have their individual beliefs or preferences, and the ``will of all'' might include individual, conflicting beliefs. The \textit{general} will, on the other hand, is some unified aggregation of the collective's will. The key component of the general will is that each person \textit{individually} is willing the object of the general will. Individuals are not just {data generators} for the general will, but rather active participants in its creation.  The dispositive component becomes not the original individual choice (or expressed preference); rather, it is the individuals' \textit{re-commitment} to the general will and their choice to accept it that gives legitimacy to the decisions arising from the general will. 

But are those decisions any good? In an echo of the discussion of correctness from Section \ref{sec:alternatives-applications}, one may wonder whether a such a procedural emphasis might come at the cost of quality outcomes. In his approach to epistemic democracy (e.g., \citet{estlund1997beyond, estlund2009democratic}), philosopher David Estlund suggests not. 
In fact, 
\blockquote[\citet{enoch2009estlund}]{

\textit{Specific political decisions in a democracy---whether correct or
incorrect---are legitimate because they are the outcomes of a democratic
procedure, and that procedure itself is legitimate because it is likely\dots to lead to correct,
that is, qualifiedly acceptable, decisions.}
}

It seems, then, that there are two components to consider when analyzing the way in which ``values-aligned'' AI functions as an aggregator of individual wills: first, active, continuous participation (after Rousseau), and second, the epistemic qualities of the aggregation process (after Estlund).

\paragraph{How might we analyze process in practice?}
For Estlund's epistemic question, an obvious starting point is in social choice, which examines mechanisms by which expressions of individual will (e.g. votes) can be aggregated, and the limitations of such methods.\footnote{See \citet{brandt2016handbook} for a textbook treatment of computational approaches.} In fact, one common argument (e.g. \citet{grofman1988rousseau}) is that such approaches are the operationalization of a Rousseauian general will. 
However, applications to the process of developing AI systems, however, are nontrivial, as the learning problems of social choice and (e.g.) RLHF have substantial differences \citep{siththaranjan2023distributional,dai2024sct,conitzer2024social}. Outside of social choice, \citet{feffer2023moral} show the limitations of a majority-vote for preference aggregation. 
It is here that improved models of (human) moral psychology and philosophy may be fruitful, perhaps by informing the design of algorithms that account for wider views of human agency.

LLMs can play a role as well, but primarily as as a tool. For instance, 
\citet{bakker2022fine} use language models to find areas of agreement; 
\citet{sorensen2023value} introduce a language model trained specifically to better formalize ``pluralistic values,'' and \citet{fish2023generative} use large language models to improve the representativeness of methods developed from social choice. In each of these works, the emphasis remains on eliciting the ``wills'' of human participants, rather than subsuming them entirely. 

On the other hand, Rousseau's ideal of continuous and affirmative participation seems much more difficult to realize.
Most closely related is recent work in \textit{participatory AI}, which seeks to bridge the gap between general participation and mechanisms for truly resolving disagreements. Recent surveys \citep{feffer2023preference, delgado2023participatory} emphasize the importance of treating participants as agents rather than just sources of data. Yet in practice---as highlighted by these surveys---there are often significant gaps in participants' ability to meaningfully contribute beyond the scope of ``preferences'' \citep{robertson2020if}, much less re-commit to the outcome of the process.

\paragraph{Beyond inputs to the learning problem} The elephant in the room behind the proposition of applying Rousseau's standard is that the political dimensions of AI (and the harms it may cause) are not restricted to the way in which preferences or values are elicited or aggregated. Rather, taking seriously the proposition of \textit{people as agents} also includes settings where people \textit{use} AI as a means to an end---which is how concrete harms are experienced today. Take, for example, the individual incidents captured in \citet{mcgregor2021preventing}. As just one example, replacing translators resulted in consequential failures in asylum cases \citep{bhuiyan2023lost}. This, too, can be understood as the outcome of a political process: what were the decisions made along the way, and by whom, that made this possible? Are there ways to expand the power of those who are subjected to AI? These questions go beyond not only the idea of ``AI as agent,'' but also beyond the idea of improving the process of building the model; this direction deserves far more consideration than we can provide here, and we leave further analysis to future work. 

\section{A Final Note}
\label{sec:discussion}

The goal of this work is not to argue, necessarily, for either interpretation of human agency as “ground truth”---after all, this is a centuries-old debate hardly settled in its originating field. As computer scientists, we understandably feel the impulse to default towards empiricism and therefore the mechanistic view---but we do suggest that readers reflect on how they understand agency in their own lives. Regardless of which view rings more personally true, we hope that this work provides new frames for thinking about AI and its associated ethical considerations, and serves to enrich the conversation about the ways in which (future iterations of) AI can---and cannot---be compared to humans.

\newpage
\section*{Acknowledgements}
We are grateful to Daniela Cammack for conversations that helped refine the argument, for suggesting references to relevant literature, and for providing feedback on early drafts.
This manuscript also benefited from feedback from Deborah Raji, Paula Gradu, Mihaela Curmei, Holly Jackson, Ben Recht, Zoë Bell, Chris Archer, and anonymous ICML reviewers. The term ``volitional'' is due to a suggestion by anonymous reviewer LoTW at ICML (replacing our previous term ``active'' agency). An earlier version of this work was presented at the AI meets Moral Psychology and Philosophy workshop at NeurIPS 2023. JD is supported by an NSF GRFP and the AI Policy Hub at UC Berkeley. 
\bibliographystyle{plainnat}
\bibliography{refs}

\newpage

\end{document}